\documentclass[prd,onecolumn]{revtex4}
\usepackage{amssymb}

\usepackage{epsfig}
\usepackage{amsmath}
\usepackage{verbatim}

\input{tcilatex}

\begin{document}

\title{Holographic glueball structure}
\author{Hilmar Forkel}
\affiliation{Departamento de F\'isica, ITA-CTA, 12.228-900 S\~ao Jos\'e dos Campos, S\~ao
Paulo, Brazil}
\affiliation{Institut f\"ur Theoretische Physik, Universit\"at Heidelberg, D-69120
Heidelberg, Germany}

\begin{abstract}
We derive and systematically analyze scalar glueball correlation functions
in both the hard-wall and dilaton soft-wall approximations to holographic
QCD. The dynamical content of the holographic correlators is uncovered by
examining their spectral density and by relating them to the operator
product expansion, a dilatational low-energy theorem and a recently
suggested two-dimensional power correction associated with the
short-distance behavior of the heavy-quark potential. This approach provides
holographic estimates for the three lowest-dimensional gluon condensates or
alternatively their Wilson coefficients, the two leading moments of the
instanton size distribution in the QCD vacuum and an effective UV gluon
mass. A remarkable complementarity between the nonperturbative physics of
the hard- and soft-wall correlators emerges, and their ability to describe
detailed QCD results can be assessed quantitatively. We further provide the
first holographic estimates for the decay constants of the $0^{++}$ glueball
and its excitations. The hard-wall background turns out to encode more of
the relevant QCD physics, and its prediction $f_{S}\simeq 0.8-0.9$ GeV for
the phenomenologically important ground state decay constant agrees inside
errors with recent QCD sum rule and lattice results.
\end{abstract}

\maketitle
\preprint{IFT-P.033/2005}

\section{Introduction}

\label{intro}

Despite more than three decades of intense experimental and theoretical
scrutiny the long predicted glueball states \cite{gel72} of Quantum
Chromodynamics (QCD) remain stubbornly elusive \cite{GbRew,pdg06}. The slow
pace of theoretical progress reflects the extraordinary complexity of the
infrared Yang-Mills dynamics which generates both the gluonic bound states
and their mixing with quarkonia. New analytical approaches for dealing with
strongly coupled gauge theories, as they have recently emerged from
gauge/gravity generalizations of the AdS/CFT correspondence \cite%
{mal97,gub98}, should therefore find rewarding and much needed applications
in the glueball sector.

Until now such applications have focused on the glueball mass spectra, which
were among the first holographically calculated observables in a variety of
more or less QCD-like gauge theories \cite{csa99} (for a review and current
developments see e.g. Refs. \cite{cac05}). More recently, glueball spectra
were also obtained in the first bottom-up \footnote{%
The bottom-up search program for the holographic QCD dual, guided by
experimental information from the gauge theory side, is often referred to as
AdS/QCD. For additional recent work in this direction see \cite%
{pol02,erl05,kat06,bro06,kar06,hir06,and06,for07,gri07,gri207,bu} and
references therein. For approaches more directly guided by the underlying,
ten-dimensional brane anatomy of the gravity dual see for example \cite%
{td,bar,tdhscatt}.} proposals for the holographic QCD dual \cite%
{bos03,det05,bos06,col07} as well as in back-reacted models \cite%
{csa06,gur07}.{}

In the present paper we are going to extend the holographic analysis of
glueball properties beyond the spectrum, by focusing on the gauge physics
content of the glueball correlation function and its spectral density. We
will relate the holographic predictions to QCD information from the operator
product expansion (OPE), a low-energy theorem based on the anomalous Ward
Identity for the dilatation current, and a recently advocated, effective UV
gluon mass. Our calculations will be based on two alternative AdS/QCD
backgrounds, namely the AdS$_{5}$ geometry with a ``hard wall'' IR brane
cutoff (of Randall-Sundrum type \cite{ran99}) in the fifth dimension \cite%
{pol02} and the dilaton-induced soft wall \cite{kar06}, which both proved
phenomenologically successful in the meson sector \cite{erl05,kar06,det05}.

A second major objective will be to provide the first holographic estimates
for the decay constants of the scalar glueball and its excitations, i.e. for
the glueball-to-vacuum matrix elements of the lowest-dimensional gluonic QCD
interpolator. These on-shell observables are of particular interest because
they contain fundamental information on glueball structure and govern the
spacial extent of the glueball (Bethe-Salpeter) wave functions. Lattice
indications for an exceptionally small size of the lowest-lying scalar
glueball \cite{def92}, for example, should translate into an unusually large
value of its decay constant. Evidence for such an enhancement was indeed
found in instanton vacuum models \cite{sch95} as well as in those QCD sum
rule analyses which include instanton contributions to the OPE coefficients %
\cite{for01,for05}.

The decay constants, which are the first glueball observables besides the
low-lying spectra for which direct (quenched) lattice results are now
available \cite{che06}, also play a crucial role in the theoretical analysis
of glueball production and decay rates. For this reason, their accurate
prediction will be instrumental in eventually meeting the two longstanding
challenges of glueball physics, i.e. the establishment of unambiguous
glueball signatures and their experimental identification. As a case in
point, the decay constants provide critical nonperturbative input for the
calculation of glueball production amplitudes in the ``gluon-rich''
radiative heavy-quarkonium decays which are currently measured at BES \cite%
{hon207}.

The paper is structured as follows:\ in section \ref{holdyn} we define the
dual bulk dynamics on which our study will be based, and we derive general
expressions for the scalar glueball correlator and the decay constants in
IR-deformed AdS$_{5}$ duals with a nontrivial dilaton background. In section %
\ref{gcor} we focus on the two AdS/QCD backgrounds mentioned above (i.e.
hard and soft wall) and derive exact analytical expressions for the
corresponding correlators and their spectral functions. We then analyze the
results by confronting them with the OPE of the QCD correlator (including
nonperturbative contributions to the Wilson coefficients), the dilatational
low-energy theorem which governs its low-momentum behavior, and the
contributions of an effective UV gluon mass. This strategy will allow us to
obtain holographic estimates for various QCD vacuum scales, i.e. three gluon
condensates (or alternatively their Wilson coefficients) and the two leading
moments of the instanton size distribution, as well as for an effective UV
gluon mass. In section \ref{gdc} we obtain holographic predictions for the
(ground and excited state) glueball decay constants and compare them to
other available theoretical results. Section \ref{sum}, finally, contains a
summary of our findings and presents our conclusions.

\section{Dual dynamics of the scalar glueball}

\label{holdyn}

The gauge/string correpondence~\cite{mal97,gub98} maps string theories in
curved, ten-dimensional spacetimes into gauges theories which live on the $d$
dimensional boundaries. For UV-conformal gauge theories like QCD with $d=4$,
the dual spacetime metric factorizes into a five-dimensional non-compact
manifold which close to its boundary approaches the anti--de Sitter space $%
\mathrm{AdS}_{5}\left( R\right) $ of curvature radius $R$, and a
five-dimensional compact Einstein space $X_{5}$ (where e.g. $%
X_{5}=S^{5}\left( R\right) $ for the maximally supersymmetric gauge theory)
with the same intrinsic size scale. The corresponding line element is~\cite%
{pol02}%
\begin{equation}
ds^{2}=g_{MN}\left( x\right) dx^{M}dx^{N}=e^{2A\left( z\right) }\frac{R^{2}}{%
z^{2}}\left( \eta _{\mu \nu }dx^{\mu }dx^{\nu }-dz^{2}\right)
+R^{2}ds_{X_{5}}^{2}  \label{metric}
\end{equation}%
(in conformal Poincar\'{e} coordinates) where $\eta _{\mu \nu }$ is the
four-dimensional Minkowski metric. Conformal invariance of the dual gauge
theory in the UV requires the absence of AdS deformations (i.e. $%
A(z)\rightarrow 0$) as $z\rightarrow 0$. Supergravity solutions suggest the
additional presence of a nontrivial dilaton background $\Phi \left( x\right) 
$, and potentially of other background fields (including Ramond-Ramond
axions, tachyons etc., see e.g. Ref. \cite{gur07,gre87}) which do, however,
not play an explicit role in the AdS/QCD duals considered below.

\subsection{Bulk action and holographic glueball correlator}

The scalar QCD glueballs are interpolated by the lowest-dimensional gluonic
operator carrying vacuum quantum numbers, 
\begin{equation}
\mathcal{O}_{S}\left( x\right) =G_{\mu \nu }^{a}\left( x\right) G^{a,\mu \nu
}\left( x\right) ,
\end{equation}%
(where $G_{\mu \nu }^{a}$ is the gluon field strength) which also figures
prominently in the anomalous dilatational Ward Identity and in the
corresponding low-energy theorems (cf. Appendix \ref{qcdsyn}). Since the
conformal dimension of $\mathcal{O}_{S}$\ is $\Delta =4$ (at the classical
level), the AdS/CFT dictionary \cite{gub98} prescribes its dual string modes 
$\varphi \left( x,z\right) $ to be the normalizable solutions of the scalar
wave equation in the bulk geometry (\ref{metric}) (and potentially other
background fields) with the UV behavior $\varphi \left( x,z\right) \overset{%
z\rightarrow 0}{\longrightarrow }z^{\Delta }\phi \left( x\right) $. The
latter implies that the square mass \footnote{%
It is straightforward to generalize the mass term to accomodate higher-twist
interpolators and to thereby describe orbital excitations of the scalar
glueball \cite{det05,bos06}.} $m_{5}^{2}R^{2}=\Delta \left( \Delta -d\right)
=0$ of the bulk field $\varphi $ vanishes, and that its minimal action has
the form%
\begin{equation}
S\left[ \varphi ;g,\Phi \right] =\frac{1}{2\kappa ^{2}}\int d^{d+1}x\sqrt{%
\left| g\right| }e^{-\Phi }g^{MN}\partial _{M}\varphi \partial _{N}\varphi
\label{sd}
\end{equation}%
(where $\kappa ^{2}$ can be related to the five-dimensional Newton constant %
\cite{gre87}) which we decompose for later use as $S=S_{M}+S_{\partial M}$
into bulk and boundary contributions with 
\begin{equation}
S_{M}\left[ \varphi ;g,\Phi \right] =-\frac{1}{2\kappa ^{2}}\int_{M}d^{d+1}x%
\sqrt{\left| g\right| }e^{-\Phi }\varphi \left[ e^{\Phi }\nabla _{M}e^{-\Phi
}g^{MN}\partial _{N}\right] \varphi  \label{sb}
\end{equation}%
($\nabla _{M}=\partial _{M}+\left| g\right| ^{-1/2}\partial _{M}\left|
g\right| ^{1/2}$) and%
\begin{equation}
S_{\partial M}\left[ \varphi ;g,\Phi \right] =\frac{1}{2\kappa ^{2}}%
\int_{\partial M}d^{d}x\left[ a^{3}\left( z\right) e^{-\Phi }\varphi
\partial _{z}\varphi \right]  \label{sds}
\end{equation}%
where $a^{2}\left( z\right) \equiv \left( R^{2}/z^{2}\right) \exp 2A\left(
z\right) $ is the warp factor. The boundary $\partial M$ consists of the UV
brane $z=\varepsilon \rightarrow 0$ and of an additional IR brane at $%
z=z_{m} $ in the hard-wall geometry.

Variation of the bulk action (\ref{sb}) with respect to $\varphi $ yields
the field equation 
\begin{equation}
e^{\Phi }\nabla _{M}e^{-\Phi }g^{MN}\partial _{N}\varphi \left( x,z\right) = 
\left[ \Delta -\left( \partial _{M}\Phi \right) g^{MN}\partial _{N}\right]
\varphi \left( x,z\right) =0  \label{fegen}
\end{equation}%
where $\Delta =\nabla _{M}\nabla ^{M}$ is the Laplace-Beltrami operator of
the metric (\ref{metric}). The action density of the solutions is finite
only on the boundary $\partial M$ while $S_{M}^{\left( \text{on-shell}%
\right) }=0$. We now specialize to dilaton fields $\Phi $ which depend
exclusively on the fifth dimension, i.e. $\Phi =\Phi \left( z\right) $. The $%
d$-dimensional Fourier transform $\hat{\varphi}\left( q,z\right) $ of the
normalizable dual modes then solves the reduced field equation%
\begin{equation}
\left[ \partial _{z}^{2}+\left( d-1\right) \left( a^{-1}\partial
_{z}a\right) \partial _{z}-\left( \partial _{z}\Phi \right) \partial
_{z}+q^{2}\right] \hat{\varphi}\left( q,z\right) =0  \label{fe}
\end{equation}%
with discrete on-shell momenta \footnote{%
The discrete spectrum is a consequence of the unbounded Sturm-Liouville
potentials for the dual modes in both hard- and soft-wall backgrounds.} $%
q^{2}=m_{n}^{2}$ in both hard- and soft-wall backgrounds. The eigenvalues $%
m_{n}^{2}$ determine the glueball mass spectrum of the boundary gauge
theory, and the corresponding orthonormalized solutions will be denoted $%
\psi _{n}\left( z\right) =N_{n}\hat{\varphi}\left( m_{n},z\right) $.

Holographic glueball correlation functions are obtained by differentiating
the bulk action of the solutions with respect to the boundary source \cite%
{gub98}. The on-shell action can be constructed with the help of the
bulk-to-boundary propagator $\hat{K}\left( q,z\right) $ \cite{gub98}, which
is the solution of the field equation (\ref{fe}) subject to the UV boundary
condition $\hat{K}\left( q;\varepsilon \rightarrow 0\right) =1$. Its
spectral representation is therefore 
\begin{equation}
\hat{K}\left( q,z\right) =-\frac{R^{3}}{\varepsilon ^{3}}\sum_{n}\frac{\psi
_{n}^{\prime }\left( \varepsilon \right) \psi _{n}\left( z\right) }{%
q^{2}-m_{n}^{2}+i\varepsilon ^{\prime }}  \label{kspec}
\end{equation}%
(where the limit $\varepsilon \rightarrow 0$ at the end of the calculation
is implied) and automatically satisfies the IR boundary condition imposed on
the $\psi _{n}\left( z\right) $. Hence one can write the solution of Eq. (%
\ref{fegen}) corresponding to a given boundary source $\varphi ^{\left(
s\right) }\left( x^{\prime }\right) $ as 
\begin{equation}
\varphi \left( x,z\right) =\int \frac{d^{4}q}{\left( 2\pi \right) ^{4}}%
e^{-iqx}\hat{K}\left( q,z\right) \int d^{4}x^{\prime }e^{iqx^{\prime
}}\varphi ^{\left( s\right) }\left( x^{\prime }\right)  \label{sln}
\end{equation}%
and obtain the associated on-shell action (which plays the role of a
generating functional) by inserting Eq. (\ref{sln}) into the surface action (%
\ref{sds}). Taking two functional derivatives with respect to $\varphi
^{\left( s\right) }$ then yields the two-point correlation function%
\begin{equation}
\left\langle T\mathcal{O}_{S}\left( x\right) \mathcal{O}_{S}\left( y\right)
\right\rangle =i\int \frac{d^{4}q}{\left( 2\pi \right) ^{4}}e^{-iq\left(
x-y\right) }\hat{\Pi}\left( -q^{2}\right)
\end{equation}%
of the scalar glueball where%
\begin{equation}
\hat{\Pi}\left( -q^{2}\right) =-\frac{R^{3}}{\kappa ^{2}}\left[ \frac{%
e^{-\Phi \left( z\right) }}{z^{3}}\hat{K}\left( q,z\right) \partial _{z}\hat{%
K}\left( q,z\right) \right] _{z=\varepsilon \rightarrow 0}.  \label{cor}
\end{equation}%
For $\Phi $ fields which vanish at the UV boundary (as does the soft wall
dilaton considered below), furthermore, the bulk-to-boundary propagator in
the form (\ref{kspec}) generates the spectral representation 
\begin{equation}
\hat{\Pi}\left( -q^{2}\right) =-\left( \frac{R^{3}}{\kappa \varepsilon ^{3}}%
\right) ^{2}\sum_{n}\frac{\psi _{n}^{\prime }\left( \varepsilon \right) \psi
_{n}^{\prime }\left( \varepsilon \right) }{q^{2}-m_{n}^{2}+i\bar{\varepsilon}%
}=-\sum_{n}\frac{f_{n}^{2}m_{n}^{4}}{q^{2}-m_{n}^{2}+i\bar{\varepsilon}}
\label{spc}
\end{equation}%
of the correlator (where a prime denotes differentiation with respect to $z$
and (divergent) contact terms are not written explicitly). The pole residues
of Eq. (\ref{spc}) at $q^{2}=m_{n}^{2}$ define the decay constants of the $n$%
-th $0^{++}$ glueball excitation as 
\begin{equation}
f_{n}:=\frac{1}{m_{n}^{2}}\left\langle 0\left| \mathcal{O}_{S}\left(
0\right) \right| 0_{n}^{++}\right\rangle =\frac{R^{3}}{\kappa m_{n}^{2}}%
\frac{\psi _{n}^{\prime }\left( \varepsilon \right) }{\varepsilon ^{3}}.
\label{dc}
\end{equation}

The physical role of the decay constants as the glueball ``wave functions at
the origin'' becomes more transparent when viewing them as the coincidence
limit of the Bethe-Salpeter amplitudes 
\begin{equation}
\chi _{n}\left( x\right) =\left\langle 0\left| 2\text{tr}\left\{ G_{\mu \nu
}\left( -\frac{x}{2}\right) U\left( -\frac{x}{2},\frac{x}{2}\right) G^{\mu
\nu }\left( \frac{x}{2}\right) \right\} \right| 0_{n}^{++}\right\rangle
\end{equation}%
(where the adjoint color parallel transporter $U\left( x,y\right) $ ensures
gauge invariance and proper renormalization of the operators is understood).
A smaller glueball size implies a higher concentration of the wave function
at the origin and consequently a larger value of $f_{n}$. Since the decay
constants are on-shell observables related to the bilinear part of the bulk
action, we expect them to be reasonably well predicted by Eq. (\ref{dc})
even though the dual dynamics\ (\ref{sd}) contains operators of minimal
dimension only.

\subsection{Comments on the scalar dual dynamics}

\label{dyncom}

We have restricted the action (\ref{sd}), i.e. the dynamics of fluctuations
dual to scalar glueballs in a metric plus dilaton background, to contain
only operators with the minimal number of fields and derivatives. This is
appropriate for the AdS/QCD candidates under consideration and will permit
us to derive analytical expressions for the exact correlators. These
restrictions also entail several typical limitations of contemporary
bottom-up models, however, which we now discuss in view of their potential
impact on the glueball sector.

A first obvious limitation is the treatment of the metric as a non-dynamical
background field, reflected in the fact that the action (\ref{sd}) contains
neither the Einstein-Hilbert term nor higher-derivative corrections to it 
\footnote{%
Backreactions of bulk fields on a dynamical metric can e.g. encode dynamical
condensate effects and implement asymptotic freedom \cite{csa06,gur07}.}.
Graviton fluctuations around the bulk metric are generally neglected as well
in bottom-up duals. If included, they could have a direct bearing on the
scalar glueball dynamics since fluctuations around a non-conformal metric
would generate a scalar ``radion'' mode (related to the fifth or radial
dimension component of the graviton) as it appears e.g. in distance
fluctuations between the branes of Randall-Sundrum models \cite{ran99,csa01}
and in dynamical dilaton-gravity models (see e.g. Ref. \cite{csa06}). By
mixing with the scalar field in the bulk action (\ref{sd}) the radion would
then modify the glueball mass spectrum and the diagonalized correlator.
Since the mixing strength decreases with increasing mass of the excitation
dual to the radion (or mixed radion-scalar), the standard neglect of the
graviton dynamics e.g. in hard-wall models is equivalent to the tacit
assumption that an \textit{a priori} unspecified stabilization mechanism
pushes this mass up far enough for radion admixtures to become negligible
(ideally by breaking conformal symmetry according to the QCD trace anomaly).

A more realistic holographic description of the glueball sector would
probably require to include further operators, potentially of higher
dimension, into the action (\ref{sd}). Those may contain additional bulk
fields, with promising candidates including spin-zero background fields
encoding condensates of relevant QCD operators and flavor-carrying gauge
fields \cite{erl05} for the description of quarkonium-gluonium mixing
effects and for specific\ glueball decay channels (see e.g. Ref. \cite{has07}%
). Operators containing a higher number of derivatives are potentially
important as well. They typically arise from stringy $\alpha ^{\prime }$
corrections in bulk regions where the curvature radius $R$ of the geometry
becomes comparable to the string length $l_{s}$. In holographic duals of
large-$N_{c}$ gauge theories such regions are expected to describe the UV
regime where the 't Hooft coupling $\lambda =g_{\text{YM}}^{2}N_{c}$ becomes
small, i.e. where $\lambda \sim \left( R^{2}/\alpha ^{\prime }\right)
^{2}\lesssim 1$ \footnote{%
High-curvature regions are further required in many dual backgrounds to
parametrically decouple unwanted Kaluza-Klein modes by pushing their mass
sufficiently far beyond the low-lying gauge theory spectrum.}. The\ lack of
asymptotic freedom in current bottom-up duals (as well as in supergravity
approximations), i.e. the fact that AdS/QCD models remain strongly coupled
in the UV\ (although they approach a conformal fixed point) \footnote{%
Several options for improvements of the UV description have recently been
explored. First indications for an improved holographic phenomenology due to
higher-dimensional operators (including higher-derivative interactions)
emerged in Refs. \cite{bas07,gri307,kim08}. Asymptotic freedom can be
implemented (at least approximately) by additional operators of stringy
origin, e.g. in the form of a dilaton potential which encodes information on
the (perturbative) QCD beta function \cite{csa06,gur07}. Another option is
to avoid a bulk gravity description of the short-distance gauge physics by
imposing an UV cutoff on the fifth dimension close to the UV brane \cite%
{eva05}.}, is therefore closely related to the absence of higher-dimensional
operators.

This discussion indicates that improvements of the AdS/QCD approach will
depend in no small measure on whether a quantitative understanding for the
impact of the strongly coupled UV regime on holographic predictions can be
developed. In the present paper we propose a strategy towards clarifying
this issue which is based on the comparison of holographic model\
predictions for hadronic correlation functions with the QCD operator product
expansion (OPE). The OPE lends itself particularly well to a systematic
diagnosis of the UV sector since it factorizes gauge theory amplitudes into
short-distance mode contributions to the Wilson coefficients and
long-distance physics contributions to local operators. We are going to
exploit this factorization property below when searching for specific traces
of the strongly coupled UV regime in the two-point function of the scalar
glueball channel and will indeed find evidence for deficiencies in the
AdS/QCD description of the perturbative Wilson coefficients (beyond the
leading conformal logarithm). Moreover, the results will suggest systematic
improvement strategies for bottom-up duals.

In view of the issues raised above one may wonder whether a decent
holographic description of asymptotically free Yang-Mills theories at large $%
N_{c}$ could at all be achieved on the basis of a \emph{local}
five-dimensional action (which may include a few higher-dimensional
operators). Fortunately, there are several indications for an affirmative
answer. A general argument due to Witten implies that the locality of the
five-dimensional bulk dynamics is ensured by the large-$N_{c}$ limit \cite%
{witjhs}. The bulk action may then be viewed as an effective string \emph{%
field} theory which contains an elementary field for each string excitation
(including those of arbitrarily high spin) while higher-dimensional
operators are suppressed by powers of $1/N_{c}$ \cite{kar06}. Moreover, $%
\alpha ^{\prime }$ corrections may be partially resummed e.g. into the
dilaton potential \cite{gur07}, and extensive QCD sum-rule \cite{qcdsr}
analyses have shown that already a few leading OPE power corrections, and
hence hopefully the few corresponding light bulk fields with controllably
small $\alpha ^{\prime }$ corrections, can capture at least the essential
properties of most hadronic ground states.

\section{Holographic glueball correlators}

\label{gcor}

The expressions derived above hold for all geometries of the form (\ref%
{metric}) and for general dilaton backgrounds $\Phi \left( z\right) $ with $%
\Phi \left( 0\right) =0$. In order to gain dynamical insight into the
holographic glueball correlator and to obtain quantitative estimates for the
decay constants, we will now consider two specific AdS/QCD backgrounds, i.e.
the AdS$_{5}$ slice of the hard IR wall geometry \cite{pol02} and the
dilaton soft wall of Ref. \cite{kar06}. In particular, we will derive
analytical expressions for the glueball correlator (\ref{cor}) and its
spectral density in the hard- and soft-wall backgrounds. Those will then be
analyzed by comparison with the QCD operator product expansion, a
dilatational low-energy theorem which governs the correlator at zero
momentum, and an effective UV gluon mass contribution of the type suggested
in Ref. \cite{che99}. The pertinent QCD information is summarized in
Appendix \ref{qcdsyn}.

\subsection{Conformal symmetry breaking by an IR brane}

\label{cbhw}

A substantial part of the successful recent AdS/QCD phenomenology (see e.g. %
\cite{pol02,erl05,bro06,gri07,det05}) was obtained on the basis of the
so-called ``hard wall'' geometry \cite{pol02}. This rather minimal
deformation of the AdS$_{5}$ metric approximately describes IR effects
including confinement by a sudden onset of conformal symmetry breaking in
the form of an IR brane at $z=z_{m}$, i.e. 
\begin{equation}
e^{2A^{\left( \text{hw}\right) }\left( z\right) }=\theta \left(
z_{m}-z\right) ,\,\ \ \ \ \ z_{m}\simeq \Lambda _{\text{QCD}}^{-1},\text{ \
\ \ \ \ }\Phi ^{\left( \text{hw}\right) }\equiv 0,  \label{hw}
\end{equation}%
which reduces the five-dimensional bulk spacetime to an AdS$_{5}$ slice.

We start our analysis of the holographic glueball correlator in this highly
symmetric background where analytical expressions are straightforward to
obtain. The bulk-to-boundary propagator $\hat{K}\left( q,z\right) $, in
particular, can be found by solving the field equation (\ref{fe}) in the
geometry (\ref{hw}), subject to the UV boundary condition $\hat{K}\left(
q;\varepsilon \right) =1$ (with $\varepsilon \rightarrow 0$) and the Neumann
IR boundary condition $\partial _{z}\hat{K}\left( q;z_{m}\right) =0$. The
result is \footnote{%
This expression exhibits the poles expected from Eq. (\ref{kspec}) at the
hard-wall glueball masses $q^{2}=m_{n}^{2}=j_{1,n}^{2}z_{m}^{-2}$ (cf. Eq. (%
\ref{mNN})).}%
\begin{equation}
\hat{K}\left( q,z\right) =\frac{\pi }{4}\left( qz\right) ^{2}\left[ \frac{%
Y_{1}\left( qz_{m}\right) }{J_{1}\left( qz_{m}\right) }J_{2}\left( qz\right)
-Y_{2}\left( qz\right) \right]  \label{b2b0}
\end{equation}%
where $J_{\nu },Y_{\nu }$ are Bessel functions and $\hat{K}\left( 0,z\right)
=1$. After plugging Eq. (\ref{b2b0}) into the general expression (\ref{cor})
and analytically continuing to spacelike momenta $Q^{2}=-q^{2}$, one ends up
with the hard-wall glueball correlator%
\begin{equation}
\hat{\Pi}\left( Q^{2}\right) =\frac{R^{3}}{8\kappa ^{2}}Q^{4}\left[ 2\frac{%
K_{1}\left( Qz_{m}\right) }{I_{1}\left( Qz_{m}\right) }-\ln \left( \frac{%
Q^{2}}{\mu ^{2}}\right) \right]  \label{hwc}
\end{equation}%
($K_{\nu },I_{\nu }$ are McDonald functions \cite{abr72}) where two contact
terms associated with UV divergent subtraction constants were discarded.

It is instructive to find the spectral density $\rho \left( s\right) $ of
the correlator (\ref{hwc}) which is defined by means of the dispersion
relation%
\begin{equation}
\hat{\Pi}\left( Q^{2}\right) =\int_{m_{1}^{2}}^{\infty }ds\frac{\rho \left(
s\right) }{s+Q^{2}}  \label{drel}
\end{equation}%
(where the necessary subtraction terms are again implied but not written
explicitly)\ and can be derived, e.g., from the well-known analyticity
properties of the McDonald functions \cite{abr72} (and the causal pole
definition) as the imaginary part of $\hat{\Pi}$/$\pi $ at timelike momenta.
The result is%
\begin{equation}
\rho \left( s\right) =\frac{R^{3}}{2\kappa ^{2}z_{m}^{2}}s^{2}\sum_{n=1}^{%
\infty }\frac{\delta \left( s-m_{n}^{2}\right) }{J_{0}^{2}\left(
j_{1,n}\right) }  \label{hwsd}
\end{equation}%
where we encountered the\ hard-wall mass spectrum $m_{n}=$ $j_{1,n}/z_{m}$
(cf. Eq. (\ref{mNN})). The spectral weight (\ref{hwsd}) is non-negative, in
agreement with general principles, and consists of a sum of zero-width
poles, as expected at large $N_{c}$ where the (infinitely many)\ glueballs
become stable against strong decay. The leading large-$s$ behavior of the
density (\ref{hwsd}) necessitates subtractions in Eq. (\ref{drel}) and
ensures the leading logarithmic $Q^{2}$ dependence of the correlator (\ref%
{hwc}).

The holographic result (\ref{hwc}) can be compared to the QCD short-distance
expansion (\ref{sde}) for $Q\gg \mu ^{2}>z_{m}^{-2}$. A standard procedure %
\cite{erl05} for fixing the overall normalization $R^{3}/\kappa ^{2}$, in
particular, is to match the coefficients of the leading conformal logarithm
in Eqs. (\ref{ope}) and (\ref{hwc}), which yields%
\begin{equation}
\frac{R^{3}}{\kappa ^{2}}=\frac{2\left( N_{c}^{2}-1\right) }{\pi ^{2}}.
\label{est}
\end{equation}%
(For a discussion of the accuracy of such estimates see Ref. \cite{eva07}.)
Below we will specialize Eq. (\ref{est}) to the phenomenologically relevant $%
N_{c}=3$ which seems - at least as far as glueball properties are concerned
- to be a surprisingly good approximation to large $N_{c}$ \cite{luc04}. We
now turn to the remaining term in the holographic correlator (\ref{hwc}),
which describes nonperturbative contributions of the boundary gauge theory
and becomes%
\begin{eqnarray}
\hat{\Pi}^{\left( \text{np}\right) }\left( Q^{2}\right) &\equiv &\frac{R^{3}%
}{4\kappa ^{2}}\frac{K_{1}\left( Qz_{m}\right) }{I_{1}\left( Qz_{m}\right) }%
Q^{4}  \notag \\
&&\overset{Qz_{m}\gg 1}{\longrightarrow }\frac{4}{\pi }\left[ 1+\frac{3}{4}%
\frac{1}{Qz_{m}}+O\left( \frac{1}{\left( Qz_{m}\right) ^{2}}\right) \right]
Q^{4}e^{-2Qz_{m}}  \label{nph}
\end{eqnarray}%
in the OPE limit $Q^{2}\gg \Lambda _{\text{QCD}}^{2}\sim z_{m}^{-2}$. Eq. (%
\ref{nph}) reveals that the hard-wall glueball correlator contains no power
corrections and that all of its nonperturbative content has an exponential $%
Q^{2}$ dependence (times powers of $Q^{2}$). In the OPE (\ref{sde}) this
exponential behavior originates from small-size instanton contributions to
the Wilson coefficients. Indeed, for $Q\gg \bar{\rho}^{-1}$ the direct
instanton contribution (\ref{ia}) becomes \textbf{\ }%
\begin{equation}
\hat{\Pi}^{\left( I+\bar{I}\right) }\left( Q^{2}\right) \overset{Q\bar{\rho}%
\gg 1}{\longrightarrow }2^{4}5^{2}\pi \zeta \bar{n}\left( Q\bar{\rho}\right)
^{3}e^{-2Q\bar{\rho}}  \label{npi}
\end{equation}%
which has exactly the momentum dependence of the first subleading term in
the non-perturbative hard-wall correlator (\ref{nph}). As shown in instanton
vacuum models \cite{sch95} and directly from the IOPE in QCD sum rules \cite%
{for01,for05}, these instanton-induced correlations are attractive and of
relatively short range $\sim \bar{\rho}$. Hence they reduce the mass and
size of the scalar glueball while increasing its decay constant.

For a quantitative comparison of holographic and instanton-induced
contributions one may approximately equate Eq. (\ref{npi}) with the second
term in Eq. (\ref{nph}). This yields the expressions \textbf{\ }%
\begin{equation}
\bar{\rho}\simeq z_{m}\text{, \ \ \ \ \ \ }\bar{n}\simeq \frac{3}{%
2^{4}5^{2}\pi ^{2}\zeta }\frac{1}{z_{m}^{4}},  \label{holest}
\end{equation}%
for the average instanton size $\bar{\rho}$ and the overall instanton
density $\bar{n}$ in terms of the hard-wall IR scale $z_{m}$. The relation $%
\bar{\rho}\simeq z_{m}$ is consistent with the duality between gauge-theory
instantons of size $\rho $ and pointlike bulk objects ($D$ instantons or $%
D\left( -1\right) $ branes in the supersymmetric case \cite{dor02})
localized at a distance $z=\rho $ from the UV boundary. However, it also
identifies the instanton's average size $\bar{\rho}$ with the maximal size $%
z_{m}$ in the AdS$_{5}$ slice, which is likely to result in an overestimate$%
. $ Indeed, the standard identification $z_{m}^{-1}\sim \Lambda _{\text{QCD}%
}\simeq 0.33$ GeV would imply $\bar{\rho}\sim 0.6$ fm, i.e. almost twice the
ILM\ value $\bar{\rho}_{\text{ILM}}\sim 0.33$ fm \cite{sch98}. As a
consequence, $\bar{n}_{\text{ILM}}\simeq 0.5$ fm$^{-4}$ \cite{sch98} would
be underestimated by the second relation in Eq. (\ref{holest}).

Besides other likely limitations of the hard-wall background including the
strongly coupled UV dynamics, the large estimate for $\bar{\rho}$ may also
reflect the absence of fundamental quark flavors in the simple dual dynamics
(\ref{sd}). The relations (\ref{holest}) (as well as other results below)
may therefore apply more accurately to pure Yang-Mills theory for which
several lattice studies indeed find larger average instanton sizes $\bar{\rho%
}\simeq 0.4-0.5$ fm \cite{gar99}. In any case, one would not expect the
instanton scales of the QCD vacuum to be precisely encoded in the hard-wall
approximation. In fact, it seems remarkable that\ this minimal background
can even semi-quantitatively reproduce the key instanton contribution to the
short-distance expansion. For a fully quantitative study of such corrections
one should resort to top-down gravity duals in which the relation between
bulk and boundary instantons can be traced exactly \cite{dor02}. Such
investigations may also shed light on the interpretation of the leading
exponential contribution to Eq. (\ref{nph}) in terms of gauge theory physics.

It is interesting to confront the hard-wall correlator with the QCD
low-energy theorem (\ref{let}). The correlator (\ref{hwc}) vanishes at $%
Q^{2}=0$ since the removal of the contact terms amounts to subtractions at $%
Q^{2}=0$. Even the contact terms do not contain a finite (or infinite)
contribution to $\hat{\Pi}\left( 0\right) $, however, and neither does the
nonperturbative part (\ref{nph}) alone which would remain after subtracting
the perturbative contributions from the spectral density, as suggested in
the original definition \cite{let}. (This is in contrast to the
one-instanton contribution (\ref{ia}) which contains a subtraction term $%
\hat{\Pi}^{\left( I+\bar{I}\right) }\left( 0\right) =2^{7}5^{2}\zeta \bar{n}$%
. The one-instanton approximation is not reliable at small $Q^{2}$, however,
where multi-instanton and other long-wavelength vacuum field contributions
are likely to dominate the correlator.) From the LET perspective this is
consistent with the absence of power corrections and gluon condensates in
the hard-wall background. As a consequence, both sides of Eq. (\ref{let})
vanish identically and the LET\ is trivially satisfied. A more complex
situation will be encountered in the soft-wall background below.

The absence of condensate effects in the hard-wall approximation is not
surprising because their purely geometrical encoding is known to require
power-law deformations \cite{hir06} of the warp factor $A\left( z\right) $
in the infrared \footnote{%
In this respect the hard-wall dual is probably closer to supersymmetric
Yang-Mills theory where the gluon condensate vanishes.}. Since large
instantons generate finite gluon condensates, this furthermore indicates
that the correlator (\ref{hwc}) receives small-size instanton contributions
only, in perfect agreement with our above discussion which indeed implies 
\textbf{\ }%
\begin{equation}
\rho ^{\left( \text{hw}\right) }\leq z_{m}\sim \mu ^{-1}
\end{equation}%
because the instanton size cannot exceed the extension of the AdS$_{5}$
slice in the fifth dimension. Hence the simple hard-wall approximation seems
to capture the fact that an essential part of the nonperturbative
contributions to the $0^{++}$ glueball correlator is hard compared to the
OPE scale $\mu $ and therefore resides in the Wilson coefficients \cite%
{for01,for05,nov280}. Since the power corrections of the OPE (\ref{ope}) are
suppressed by unusually small Wilson coefficients, furthermore, the
hard-wall background may indeed provide a reasonable first approximation to
the scalar glueball correlator.

\subsection{Dilaton-induced conformal symmetry breaking}

\label{swcor}

A well-known\ shortcoming of the hard-wall background (\ref{hw}) is that it
predicts squared hadron masses to grow quadratically with high radial, spin
and orbital excitation quantum numbers \cite{kat06,bro06,det05}, in contrast
to the linear trajectories expected from semiclassical flux tube models \cite%
{shi05}. This problem manifests itself also in the hard-wall glueball
spectra (cf. Eqs. (\ref{mND}), (\ref{mNN})) which do not reproduce the
expected linear Pomeron trajectory \cite{don02,mey05}.

The presence of a nontrivial dilaton background field $\Phi \left( z\right)
\propto z^{2}$ was recently proposed as an economical remedy for this
problem in the meson \cite{kar06} and glueball \cite{col07} sectors. In the
simplest version of the resulting gravity dual, conformal symmetry breaking
in the IR is the exclusive task of the dilaton while the geometry (\ref%
{metric}) remains undeformed AdS$_{5}$, i.e.%
\begin{equation}
A^{\left( \text{sw}\right) }\left( z\right) \equiv 0,\text{ \ \ \ \ \ }\Phi
^{\left( \text{sw}\right) }\left( z\right) =\lambda ^{2}z^{2}.  \label{sw}
\end{equation}%
In the present section we derive and analyze the scalar glueball correlator
in this ``dilaton soft-wall'' background. (Alternative holographic
realizations of linear trajectories have been obtained for mesons or
glueballs in Refs. \cite{and206,kru05,cas07,hua07} and for both mesons and
baryons in Ref. \cite{for07}.)

Although the background (\ref{sw}) is somewhat more complex than the minimal
hard-wall geometry (\ref{hw}), one can still find a closed integral
representation for the corresponding scalar bulk-to-boundary propagator (\ref%
{kspec}),%
\begin{equation}
\hat{K}\left( q;z\right) =\frac{q^{2}}{4\lambda ^{2}}\left( \frac{q^{2}}{%
4\lambda ^{2}}-1\right) \int_{0}^{1}dx\left( 1-x\right) x^{-\left[
q^{2}/\left( 4\lambda ^{2}\right) +1\right] }e^{-\frac{x}{1-x}\lambda
^{2}z^{2}},  \label{ksw}
\end{equation}%
which may be rewritten in terms of confluent hypergeometric functions. Eq. (%
\ref{ksw}) is easily shown to be the solution of the field equation (\ref{fe}%
) in the background (\ref{sw}) which satisfies the UV boundary condition $%
\hat{K}\left( q;0\right) =1$ and additionally $\hat{K}\left( 0;z\right) =1$.
Inserting the expression (\ref{ksw}) into Eq. (\ref{cor}) leads to 
\begin{equation}
\hat{\Pi}\left( Q^{2}\right) =-2\frac{R^{3}\lambda ^{2}}{\kappa ^{2}}\frac{%
Q^{2}}{4\lambda ^{2}}\left( \frac{Q^{2}}{4\lambda ^{2}}+1\right)
\lim_{\varepsilon \rightarrow 0}\frac{1}{\varepsilon ^{2}}%
\int_{0}^{1}dxx^{Q^{2}/\left( 4\lambda ^{2}\right) }e^{-\frac{x}{1-x}\lambda
^{2}\varepsilon ^{2}}
\end{equation}%
which is the exact soft-wall correlator at spacelike momenta $q^{2}=-Q^{2}$.
The remaining integral can be performed analytically. This is conveniently
done by absorbing the small-$\varepsilon $ singularity into the integrand
such that the branch cut structure becomes manifest. One then obtains%
\begin{equation}
\hat{\Pi}\left( Q^{2}\right) =-\frac{2R^{3}\lambda ^{4}}{\kappa ^{2}}\frac{%
Q^{2}}{4\lambda ^{2}}\left( \frac{Q^{2}}{4\lambda ^{2}}+1\right) \Gamma
\left( \frac{Q^{2}}{4\lambda ^{2}}+1\right) \lim_{\varepsilon \rightarrow
0}U\left( \frac{Q^{2}}{4\lambda ^{2}}+2,2,\lambda ^{2}\varepsilon ^{2}\right)
\end{equation}%
where $U\left( a,b,z\right) $ is the (multi-valued)\ confluent
hypergeometric function \cite{abr72}. After taking the $\varepsilon
\rightarrow 0$ limit and discarding two divergent contact terms, one finally
ends up with%
\begin{equation}
\hat{\Pi}\left( Q^{2}\right) =-\frac{2R^{3}}{\kappa ^{2}}\lambda ^{4}\left[
1+\frac{Q^{2}}{4\lambda ^{2}}\left( 1+\frac{Q^{2}}{4\lambda ^{2}}\right)
\psi \left( \frac{Q^{2}}{4\lambda ^{2}}\right) \right]  \label{swc}
\end{equation}%
in terms of the digamma function $\psi \left( z\right) =\Gamma ^{\prime
}\left( z\right) /\Gamma \left( z\right) $ \cite{abr72}.

As in the hard-wall case, we begin our analysis of the correlator (\ref{swc}%
) by deriving its spectral density from the dispersion relation (\ref{drel})
(where the lower boundary of the integration region is now $s_{\min
}=m_{0}^{2}$, see below)\ as the imaginary part of $\hat{\Pi}/\pi $ at
timelike momenta. The analyticity structure of the digamma function \cite%
{abr72} and the causal pole definition then imply%
\begin{equation}
\rho \left( s\right) =\frac{\lambda ^{2}R^{3}}{2\kappa ^{2}}s\left(
s-m_{0}^{2}/2\right) \sum_{n=0}^{\infty }\delta \left( s-m_{n}^{2}\right) .
\label{swsd}
\end{equation}%
The spectral density (\ref{swsd}) is non-negative for $s\geq m_{0}^{2}/2$
and consists, as it's hard-wall counterpart (\ref{hwsd}) and as expected at
large $N_{c}$, of a sum of zero-width poles at the soft-wall masses $%
m_{n}^{2}=4\left( n+2\right) \lambda ^{2}$ (cf. Eq. (\ref{msw})). The
leading large-$s$ behavior again encodes the conformal large-$Q^{2}$
behavior of the correlator.

In order to compare the holographic soft-wall correlator to the OPE (\ref%
{sde}) at $Q^{2}\gg \Lambda _{\text{QCD}}^{2}$, we rewrite Eq. (\ref{swc})
for $Q^{2}\gg 4\lambda ^{2}$ by means of the asymptotic expansion for the
digamma function \cite{abr72} and the Bernoulli numbers $B_{2n}=\left(
-1\right) ^{n-1}2\left( 2n\right) !\zeta \left( 2n\right) /\left( 2\pi
\right) ^{2n}$ ($\zeta \left( z\right) $ is Riemann's zeta function) as 
\begin{eqnarray}
\hat{\Pi}\left( Q^{2}\right) &=&-\frac{2R^{3}}{\kappa ^{2}}\lambda ^{4}\left[
1+\frac{Q^{2}}{4\lambda ^{2}}\left( 1+\frac{Q^{2}}{4\lambda ^{2}}\right)
\left( \ln \frac{Q^{2}}{4\lambda ^{2}}-\frac{2\lambda ^{2}}{Q^{2}}%
-\sum_{n=1}^{\infty }\frac{B_{2n}}{2n}\left( \frac{4\lambda ^{2}}{Q^{2}}%
\right) ^{2n}\right) \right]  \notag \\
&=&-\frac{2}{\pi ^{2}}Q^{4}\left[ \ln \frac{Q^{2}}{\mu ^{2}}+\frac{4\lambda
^{2}}{Q^{2}}\ln \frac{Q^{2}}{\mu ^{2}}+\frac{2^{2}5}{3}\frac{\lambda ^{4}}{%
Q^{4}}-\frac{2^{4}}{3}\frac{\lambda ^{6}}{Q^{6}}+\frac{2^{5}}{15}\frac{%
\lambda ^{8}}{Q^{8}}+...\right] .  \label{c}
\end{eqnarray}%
(In the last line we have adapted the correlator to the OPE scale $\mu $ by
absorbing additional, finite pieces into the contact terms). Note that the
expansion coefficients grow factorially with the power of $\lambda
^{2}/Q^{2} $, as expected from QCD. The coefficients of the conformal
logarithm in Eq. (\ref{c}) and in the hard-wall correlator (\ref{hwc}) are
identical. This is because large momenta $Q$ probe the $z\rightarrow 0$
region where neither the dilaton nor the IR brane affect the correlator, so
that the same AdS$_{5} $-induced logarithm governs its behavior in both
hard- and soft-wall backgrounds. Hence comparison with the perturbative
gluon loop of the OPE (\ref{ope}) fixes the normalization $R^{3}/\kappa ^{2}$
as in Eq. (\ref{est}) and as anticipated in the second line of Eq. (\ref{c}).

In addition to the leading conformal logarithm, the expansion (\ref{c})
contains an infinite tower of power corrections. Comparison with the OPE (%
\ref{ope}) suggests them to be related to the gauge-theory condensates 
\begin{equation}
\left\langle \mathcal{O}_{D}\right\rangle \sim \lambda ^{D}\sim \Lambda _{%
\text{QCD}}^{D}  \label{cs}
\end{equation}%
of $D=4,6,8$, ... dimensional (local, gauge-invariant) composite operators.
The appearance of the scale factor $\lambda ^{D}$ shows that the soft-wall
power corrections are entirely dilaton-induced, in contrast to those arising
from (hadron channel dependent) deformations of the metric in the geometric
approach \cite{hir06} or from additional scalar background fields.
Tentatively equating the coefficients of the $D=4,6$ and $8$ terms (without $%
O\left( \alpha _{s}\right) $ corrections) to their OPE counterparts in Eq. (%
\ref{ope}) allows for a more quantitative check of the holographic expansion
(\ref{c}). The resulting relations for the three lowest-dimensional gluon
condensates (defined at the OPE scale $\mu \sim $ 1 GeV) are%
\begin{align}
\left\langle G^{2}\right\rangle & \simeq -\frac{10}{3\pi ^{2}}\lambda ^{4},
\label{g2} \\
\left\langle gG^{3}\right\rangle & \simeq \frac{4}{3\pi ^{2}}\lambda ^{6},
\label{g3} \\
\left\langle G^{4}\right\rangle & \simeq -\frac{8}{15\pi ^{3}\alpha _{s}}%
\lambda ^{8}.  \label{g4}
\end{align}%
These holographic estimates indeed reproduce the order of magnitude expected
from QCD. This is mostly because their scale is set by the dilaton IR
parameter $\lambda \sim \sqrt{2}\Lambda _{\text{QCD}}$ \cite{for07} which
generates the mass gap and because the coefficients in Eqs. (\ref{g2}) and (%
\ref{g3}) are more or less of order unity. The sign of the most reliably
determined four-dimensional QCD gluon condensate $\left\langle
G^{2}\right\rangle \sim 0.4-1.2$ GeV$^{4}$ is positive, however, in contrast
to Eq. (\ref{g2}). QCD estimates of both signs exist for the three-gluon
condensate, namely the lattice prediction $\langle gG^{3}\rangle \simeq
-1.5\,\langle \alpha _{s}G^{2}\rangle ^{3/2}$ \cite{g3lat} and the
single-instanton value $\langle gG^{3}\rangle \simeq 0.27$ $\mathrm{GeV}%
^{2}\langle \alpha _{s}G^{2}\rangle $. The signs of Eqs. (\ref{g2}) and (\ref%
{g4}), furthermore, are at odds with the factorization approximation \cite%
{nov279}%
\begin{equation}
\left\langle G^{4}\right\rangle \simeq \frac{9}{16}\langle G^{2}\rangle ^{2}
\end{equation}%
for the four-gluon condensate combination (\ref{g4fact}). These shortcomings
indicate that the tentative adoption of the (leading-order) perturbative QCD
Wilson coefficients for the analysis of the holographic power corrections (%
\ref{c}) is questionable. It will be revised on more physical grounds below.
(Recall, furthermore, that the scalar background field of the soft wall (\ref%
{sw}) does not correspond to a $\Delta =4$ operator.)

In addition to the OPE-type power corrections of Eq. (\ref{ope}), the
holographic soft-wall correlator (\ref{c}) contains a two-dimensional power
correction (times a logarithm) which cannot appear in the OPE since QCD
lacks a corresponding (gauge-invariant and local) composite operator.
However, a two-dimensional power correction of exactly this type\ was
advocated some time ago and argued to improve QCD sum rule results in
several hadron channels \cite{che99}. More specifically, when (possibly
renormalon-related) linear contributions to the heavy-quark potential at 
\emph{short} distances are approximately accounted for by an effective gluon
mass $\bar{\lambda}$, the latter produces the correction \cite{che99}%
\begin{equation}
\hat{\Pi}^{(\text{CNZ})}\left( Q^{2}\right) =-\frac{2}{\pi ^{2}}Q^{4}\ln 
\frac{Q^{2}}{\mu ^{2}}\left( 1+6\frac{\bar{\lambda}^{2}}{Q^{2}}+...\right)
\label{cnz}
\end{equation}%
to the leading logarithm of the glueball correlator which has precisely the
form of the second term in Eq. (\ref{c}). The appearance of this term
supports previous arguments which tentatively relate the quadratic behavior
of the soft-wall dilaton background field (\ref{sw}) or alternatively of $%
A\left( z\right) $ \cite{and06,for07,csa06,and206} to a two-dimensional
power correction and possibly to a two-dimensional nonlocal gluon condensate %
\cite{gub01}. Comparison of the $\bar{\lambda}^{2}$ correction in Eq. (\ref%
{cnz}) with its counterpart in Eq. (\ref{c}) leads to the holographic
estimate%
\begin{equation}
\bar{\lambda}^{2}\simeq \frac{2}{3}\lambda ^{2}  \label{uvgm}
\end{equation}%
and with the approximate identification $\lambda \simeq \sqrt{2}\Lambda _{%
\text{QCD}}$ further to $\bar{\lambda}^{2}\simeq 0.15$ GeV$^{2}$ which is
indeed of the expected magnitude \cite{che99}. However, as in the case of
the leading OPE power corrections the sign turns out to be opposite to QCD
expectations, i.e. the square mass (\ref{uvgm}) is not tachyonic.

The complete reproduction of the $Q^{2}$ dependence contained in the QCD
short-distance expansion (to leading order in $\alpha _{s}$) by the
soft-wall dynamics, albeit with the signs of at least the leading power
corrections opposite to QCD expectations, suggests an interpretation which
may help to disentangle the holographic predictions for Wilson coefficients
and condensates even though they appear as products in the power
corrections. Indeed, the dimensions of the condensates are generated by the
operators of the OPE which in turn are renormalized at relatively small
scales $\mu \sim 1$ GeV and hence IR\ dominated. This makes it likely that
the condensate part of the OPE and consequently the form of the power
corrections and their scaling behavior are better reproduced by the
strong-coupling dynamics of the soft-wall model, and that the deviations
from the\ OPE should reside mainly in the Wilson coefficients (cf. Sec. \ref%
{dyncom}). The lack of perturbative $Q^{2}$-dependence due to radiative $%
O\left( \alpha _{s}\right) $ corrections (cf. App. \ref{qcdsyn}) in the
soft-wall correlator provides additional support for this interpretation. It
could be further tested by extending our comparison of holographic
correlators with the OPE to other hadron channels. Indeed, since the
condensates are universal (i.e. channel independent) while the Wilson
coefficients are not, one would expect inconsistent soft-wall condensate
predictions in different hadron correlator channels when relying on the
questionable assumption that the soft wall dynamics approximates their
Wilson coefficients.

Tentatively assuming that the soft-wall dynamics approximately reproduces
the values of the QCD (or Yang-Mills)\ condensates, on the other hand, one
may obtain holographic estimates for the Wilson coefficients. The soft-wall
prediction for the (leading-order) perturbative gluon condensate coefficient 
$C_{\left\langle G^{2}\right\rangle }^{\left( \text{QCD,lo}\right) }\equiv
B_{0}$, e.g., becomes with $\left\langle G^{2}\right\rangle \simeq \left(
20/3\right) \Lambda _{\text{QCD}}^{4}$ \cite{for05} and $\lambda \simeq 
\sqrt{2}\Lambda _{\text{QCD}}$, $\Lambda _{\text{QCD}}\simeq 0.33$ GeV \cite%
{pdg06} 
\begin{equation}
C_{\left\langle G^{2}\right\rangle }^{\left( \text{sw}\right) }\simeq -\frac{%
8}{\pi ^{2}}=-\frac{2}{\pi ^{2}}C_{\left\langle G^{2}\right\rangle }^{\left( 
\text{QCD,lo}\right) }.  \label{CG2}
\end{equation}%
This prediction is of smaller size than the QCD value and has the opposite
sign. As discussed above, it is suggestive to attribute at least part of
these discrepancies to the strongly-coupled UV regime of the soft-wall
model, although the estimate (\ref{CG2}) is prone to additional error
sources including the current uncertainties in the QCD value of the gluon
condensate and its sensitivity to the presence of light quark flavors. The
uncertainties in the analogous predictions for the Wilson coefficients of
higher-dimensional operators would be further increased by the less reliably
known QCD values of the corresponding condensates. One should note, finally,
that the above approximate separation of hard and soft (i.e. $k\gtrless \mu $%
) contributions to the holographic predictions would not work for the gluon
mass term since both the mass $\bar{\lambda}$ and its coefficient receive UV
contributions.

The soft-wall correlator in its subtracted form (\ref{swc}) fails to satisfy
the low-energy theorem (\ref{let}): Eq. (\ref{g2}) (if taken literally)\
implies a finite RHS while Eq. (\ref{swc}) gives $\hat{\Pi}\left( 0\right)
=0 $ (even before discarding the contact terms), i.e. a vanishing LHS. Of
course this comparison should be considered naive since contact terms are
renormalization scheme dependent and devoid of intrinsic physical meaning,
but other subtraction procedures, including the subtraction of the conformal
logarithm suggested in the original LET definition \cite{let}, would lead to
the same result. In fact, the simple soft-wall background does not correctly
represent the physics of the QCD trace anomaly on which the LET (\ref{let})
is based: the AdS$_{5}$ metric (which is dual to the energy-momentum tensor $%
T_{\mu \nu }$ of the gauge theory \cite{bia01} on the flat boundary) implies 
$\left\langle T_{\mu }^{\mu }\right\rangle _{\text{metric}}=0$ since the AdS$%
_{5}$ Weyl anomaly vanishes \cite{hen98}, and there is no scalar background
dual to the $\Delta =4$ gluon condensate operator which appears on the RHS
of the LET and in the matter anomaly contribution to $\left\langle T_{\mu
}^{\mu }\right\rangle $. (The soft-wall dilaton would naively correspond to
a local $\Delta =2$ operator which does not exist in QCD but arises in (e.g.
effective dual color \cite{bak03}) theories with spontaneously broken gauge
symmetry.)

To summarize, it is remarkable that the soft-wall background reproduces all
qualitative features of the short-distance QCD correlator, i.e. exactly
those powers and logarithms which appear in QCD, and even the hypothetical
logarithmic corrections due to an UV gluon mass. The signs (and sizes)\ of
both leading power corrections differ from those preferred in QCD, however,
which we expect to be at least partly due to the failure of the strongly
coupled UV regime to describe the perturbative QCD Wilson coefficients.
Since QCD sum-rule analyses show that results for ground state masses and
couplings (decay constants) depend sensitively on magnitude and sign of the
leading power corrections, it is likely that the soft-wall predictions will
be contaminated by this shortcoming.

The addition of stringy corrections to the minimal bulk action (\ref{sd})
may be a promising direction for improving the soft-wall description in the
UV. Indeed, first attempts to allow for such higher-dimensional operators in
the action of holographic models \cite{bas07,gri307,kim08} show that they
can generate substantial contributions to the power corrections. Similar
operators of stringy origin, including e.g. tachyon fields or $\alpha
^{\prime }$ corrections analogous to those considered in the vector meson
sector \cite{gri307}, can therefore be expected to improve the soft-wall
prediction for the short-distance correlator in the scalar\ glueball channel.

The comparison of the above results with those for the hard-wall correlator
in Sec. \ref{cbhw} shows that the whole nonperturbative momentum dependence
of the known IOPE (up to radiative corrections)\ is reproduced by the
holographic hard- and soft-wall correlators in a fully complementary
fashion: while the soft-wall correlator contains all OPE power corrections
of the types induced either by gluon condensates or by an effective UV gluon
mass, the nonpertubative physics in the hard-wall correlator is exponential
and includes a term which reproduces the behavior of the leading instanton
contributions. This complementarity of the nonperturbative physics
represented by both dual backgrounds is likely to persist in other hadron
correlators as well (at least at distances smaller than the inverse QCD
scale) and can be exploited for diagnostic purposes, e.g. by tracing the
impact of different parts of the gauge dynamics on hadron observables (see
below).

\section{Glueball decay constants}

\label{gdc}

In the following section we obtain quantitative holographic predictions for
the glueball decay constants (\ref{dc}) in both hard-wall and dilaton
soft-wall backgrounds and discuss the underlying physics.

\subsection{Hard wall IR brane}

The values of the glueball decay constants in the hard-wall approximation
may serve as a benchmark for the results of more elaborate holographic
duals. We calculate them directly from the normalizable solutions \cite%
{bos03,det05}%
\begin{equation}
\psi _{n}\left( z\right) =N_{n}\left( m_{n}z\right) ^{2}J_{2}\left(
m_{n}z\right)  \label{hwm}
\end{equation}%
(where $n=1,2,3,...$) of the massless field equation (\ref{fe}) in the AdS$%
_{5}$ slice (\ref{hw}), which we require to satisfy (in addition to the
AdS/CFT boundary condition $\psi _{n}\left( z\right) \rightarrow z^{\Delta }$
at $z=\varepsilon \rightarrow 0$) either Dirichlet (D) or Neumann (N)
boundary conditions on the IR brane. The normalization constants $N_{n}$ are
determined by the inner product of the eigenmodes, i.e. by requiring $%
\int_{0}^{z_{m}}dz\left( R/z\right) ^{3}\psi _{n}^{2}=1$. For Dirichlet
boundary conditions $\psi _{n}\left( z_{m}\right) =0$ one then obtains the
masses \cite{bos03,det05} and normalizations 
\begin{equation}
m_{n}^{\left( \text{D}\right) }=\frac{j_{2,n}}{z_{m}},\text{ \ \ \ \ \ }%
N_{n}^{\left( \text{D}\right) }=\frac{\sqrt{2}}{m_{n}^{\left( \text{D}%
\right) 2}R^{3/2}z_{m}\left| J_{1}\left( j_{2,n}\right) \right| }
\label{mND}
\end{equation}%
while the alternative Neumann\ boundary conditions $\psi _{n}^{\prime
}\left( z_{m}\right) =0$ yield the spectrum \cite{bos06} and normalization
constants 
\begin{equation}
m_{n}^{\left( \text{N}\right) }=\frac{j_{1,n}}{z_{m}},\text{ \ \ \ \ \ }%
N_{n}^{\left( \text{N}\right) }=\frac{\sqrt{2}}{m_{n}^{\left( \text{N}%
\right) 2}R^{3/2}z_{m}\left| J_{0}\left( j_{1,n}\right) \right| }.
\label{mNN}
\end{equation}%
Here $j_{m,n}$ denotes the $n$-th zero of the $m$-th Bessel function \cite%
{abr72}. Although the normalization constants do not affect the mass
spectra, they provide a crucial overall scale for the decay constants.

From the general expression (\ref{dc}) for the decay constants and the
hard-wall eigenmodes (\ref{hwm})\ one then finds%
\begin{equation}
f_{n}=\lim_{\varepsilon \rightarrow 0}\frac{R^{3}}{\kappa m_{n}^{2}}\frac{%
\psi _{n}^{\prime }\left( \varepsilon \right) }{\varepsilon ^{3}}=\frac{N_{n}%
}{2}\frac{R^{3}}{\kappa }m_{n}^{2}
\end{equation}%
or more specifically for the above two IR boundary conditions%
\begin{equation}
f_{n}^{\left( \text{D}\right) }=\frac{1}{\sqrt{2}\left| J_{1}\left(
j_{2,n}\right) \right| }\frac{R^{3/2}}{\kappa z_{m}},\text{ \ \ \ \ \ }%
f_{n}^{\left( \text{N}\right) }=\frac{1}{\sqrt{2}\left| J_{0}\left(
j_{1,n}\right) \right| }\frac{R^{3/2}}{\kappa z_{m}}.  \label{fhw}
\end{equation}%
The expression for $f_{n}^{\left( \text{N}\right) }$ can alternatively be
obtained by comparing the spectral density (\ref{hwsd}) of the Neumann
hard-wall correlator to the general spectral representation (\ref{spc}).
This provides a useful cross-check on our calculations.

After fixing the overall normalization factor $R^{3/2}/\kappa $ by
comparison with the QCD gluon loop contribution according to Eq. (\ref{est}%
), both masses and decay constants are given (by Eqs. (\ref{mND}), (\ref{mNN}%
) and (\ref{fhw})) in terms of only one adjustable parameter, i.e. the IR
scale $z_{m}^{-1}\sim \Lambda _{\text{QCD}}$ of the hard-wall geometry which
has to be determined from independent input. The resulting quantitative
predictions for $f_{n}$ will be discussed in Sec. \ref{quant}.

\subsection{Dilaton-induced soft wall}

In the AdS$_{5}$ -- dilaton background (\ref{sw}), the solutions of the
scalar field equation (\ref{fe}) turn into Kummer's confluent hypergeometric
functions \cite{col07}. The spectrum-generating normalizable modes then form
the subset of Kummer functions whose power series expansion truncates to a
finite polynomial which turns out to be of generalized Laguerre type $%
L_{n}^{\left( 2\right) }$ \cite{abr72}, i.e.%
\begin{equation}
\psi _{n}\left( z\right) =N_{n}\lambda ^{4}z^{4}{}_{1}F_{1}\left(
-n,3,z^{2}\lambda ^{2}\right) =N_{n}\lambda ^{4}z^{4}\frac{n!}{\left(
3\right) _{n}}L_{n}^{\left( 2\right) }\left( \lambda ^{2}z^{2}\right)
\label{swnm}
\end{equation}%
where $n=0,1,2,...$, $\left( a\right) _{n}\equiv a\left( a+1\right) \left(
a+2\right) ...\left( a+n-1\right) $ and $_{1}F_{1}$ is a confluent
hypergeometric function \cite{abr72}. The ensuing restriction to discrete
eigenvalues $q^{2}=m_{n}^{2}$ yields the glueball mass spectrum \cite{col07}%
\begin{equation}
m_{n}^{2}=4\left( n+2\right) \lambda ^{2}  \label{msw}
\end{equation}%
and relates the mass gap $m_{0}=2\sqrt{2}\lambda $ to the dilaton background
scale. In contrast to its hard-wall counterparts (\ref{mND}) and (\ref{mNN}%
), the soft-wall spectrum (\ref{msw}) grows linearly with $n$ and thus
generates a Pomeron-type trajectory \cite{don02,mey05}. The normalization
constants $N_{n}$ are obtained from the inner product in the eigenmode space
by demanding 
\begin{equation}
\int_{0}^{\infty }dz\left( \frac{R}{z}\right) ^{3}e^{-\lambda ^{2}z^{2}}\psi
_{n}^{2}\left( z\right) =1
\end{equation}%
which yields%
\begin{equation}
N_{n}=\lambda ^{-1}R^{-3/2}\left( I_{n}\right) ^{-1/2}
\end{equation}%
in terms of the integrals%
\begin{equation*}
I_{n}:=\int_{0}^{\infty }d\xi e^{-\xi ^{2}}\xi ^{5}{}_{1}F_{1}^{2}\left(
-n,3,\xi ^{2}\right) =\frac{n!}{\left( 3\right) _{n}}=\frac{2}{\left(
n+1\right) \left( n+2\right) }.
\end{equation*}%
(Note that $N_{n}\propto \left( I_{n}\right) ^{-1/2}\rightarrow 2^{-1/2}n$
for $n\gg 3$, and to a rather good approximation already for $n\gtrsim 3$.)

From the general expression (\ref{dc}) one then obtains the glueball decay
constants in the soft-wall background as 
\begin{equation}
f_{n}^{\left( \text{sw}\right) }=4I_{n}^{-1/2}\frac{\lambda ^{3}R^{3/2}}{%
m_{n}^{2}\kappa }=\frac{1}{\sqrt{2}}\sqrt{\frac{n+1}{n+2}}\frac{\lambda
R^{3/2}}{\kappa }.{}  \label{fsw}
\end{equation}%
This expression shows that the $f_{n}^{\left( \text{sw}\right) }$ increase
by only about 40\% from $n=0$ to $n=\infty $ and approach the universal
value $f_{\infty }^{\left( \text{sw}\right) }=$ $\lambda R^{3/2}/\left( 
\sqrt{2}\kappa \right) $ towards higher excitation levels\ rather fast, in
contrast to the weak but unbounded increase of their hard-wall counterparts (%
\ref{fhw}).

Eq. (\ref{fsw}) can be checked by alternatively deriving it from the
spectral density (\ref{swsd}), and the factor $R^{3/2}/\kappa $ can again be
estimated by Eq. (\ref{est}) which continues to hold in the soft-wall
background. The dilaton scale $\lambda $ will be approximately determined in
Sec. \ref{quant}.

\subsection{Quantitative analysis}

\label{quant}

We restrict our quantitative decay-constant estimates to the glueball ground
state, i.e. to $f_{1}^{\left( \text{hw}\right) }\equiv f_{S}^{\left( \text{hw%
}\right) }$ and $f_{0}^{\left( \text{sw}\right) }\equiv f_{S}^{\left( \text{%
sw}\right) }$, since only $f_{S}$ will be of phenomenological relevance in
the foreseeable future and since independent theoretical information on it
is currently available. (The extension to higher resonances by means of
formulae (\ref{fhw}) and (\ref{fsw}) is of course immediate.) After having
fixed the correlator normalization $R^{3}/\kappa ^{2}$ according to Eq. (\ref%
{est}) in both backgrounds, it remains to\ determine the IR scale $%
z_{m}^{-1} $ ($\lambda $) of the hard- (soft-) wall gravity dual. In order
to get an idea of how the uncertainties involved in different scale-setting
approaches affect the decay constant predictions, we will discuss several
alternative possibilities.

A commonly adopted strategy for fixing the IR scale is to match the
holographic ground state mass to lattice results. Uncertainties of this
method include the still rather large scale-setting ambiguity of quenched
lattice predictions \cite{bal01} and the neglected light-quark effects
(including quarkonium mixing and decay channels) which may substantially
reduce the quenched scalar glueball masses \cite{har02}. Nevertheless, the
quenched masses can serve as a useful benchmark for scale-setting purposes,
in particular because it is not clear how far quark effects are accounted
for in the simple dual dynamics which we consider here.

We therefore base our first estimate on a typical quenched glueball mass $%
m_{S}\simeq 1.5$ GeV \cite{che06,mey05,lee00}, which coincides with the mass
of the experimental glueball candidate $f\left( 1500\right) $ and fixes the
IR scale of the Dirichlet (Neumann) hard wall at $z_{m}^{\left( \text{D}%
\right) -1}=0.29$ GeV ($z_{m}^{\left( \text{N}\right) -1}=0.39$ GeV) and
that of the soft wall at $\lambda =0.43$ GeV. (Note that the values for $%
z_{m}$ and $\lambda /\sqrt{2}$ are indeed rather close to $\Lambda _{\text{%
QCD}}$, as assumed in the qualitative estimates of Sec. \ref{gcor}.) When
inserted into Eqs. (\ref{fhw}) and (\ref{fsw}), these scales lead to the
predictions%
\begin{eqnarray}
f_{S}^{\left( \text{D}\right) } &=&0.77\text{ GeV,} \\
f_{S}^{\left( \text{N}\right) } &=&0.87\text{ GeV}
\end{eqnarray}%
in the hard-wall geometry and to the about three times smaller value%
\begin{equation}
f_{S}^{\left( \text{sw}\right) }=0.28\text{ GeV}  \label{fsw1}
\end{equation}%
in the soft-wall background. Since both of the parameters which underly
these results were fixed in the glueball sector and in the absence of quarks
(recall that the estimate (\ref{est}) is based on the free gluon loop), the
above values are perhaps best associated with pure Yang-Mills theory.

Alternatively, one can determine the value of the hard IR wall cutoff in the
classical hadron sector, e.g. from a fit to $\pi $ and $\rho $ meson
properties as in Refs. \cite{erl05,det05}. The typical result is $%
z_{m}^{-1}\simeq 0.35$ GeV and yields%
\begin{eqnarray}
f_{S}^{\left( \text{D}\right) } &=&0.93\text{ GeV,} \\
f_{S}^{\left( \text{N}\right) } &=&0.78\text{ GeV.}
\end{eqnarray}%
The corresponding ground state glueball mass predictions are then $%
m_{S}^{\left( \text{D}\right) }=1.80$ GeV and $m_{S}^{\left( \text{N}\right)
}=1.34$ GeV (where $m_{S}\equiv m_{1}$). The latter is significantly smaller
than most quenched lattice results but close to the $f\left( 1270\right) $\
and to results of K-matrix analyses of scalar resonance data \cite{ani03},
mixing schemes with only one $0^{++}$ multiplet below 1.8 GeV \cite{och03},
a topological knot model \cite{fad04} and the QCD sum rule prediction $%
m_{S}=1.25\pm 0.2$ GeV \cite{for05}. One might speculate that fixing $%
z_{m}^{-1}$ in the flavored meson sector takes some light-quark effects into
account and hence corresponds to a lower, unquenched value of the scalar
glueball mass (at least under Neumann IR boundary conditions). For an
alternative estimate of the soft-wall IR mass scale $\lambda $ (which has
not yet been determined in the meson sector), finally, one can use its
approximate relation $\lambda \simeq \sqrt{2}\Lambda _{\text{QCD}}\simeq
0.49 $ GeV (cf. e.g. Ref. \cite{for07}) to the QCD scale $\Lambda _{\text{QCD%
}}\sim 0.33$ GeV \cite{pdg06} (for three light quark flavors). This yields
the soft-wall prediction%
\begin{equation}
f_{S}^{\left( \text{sw}\right) }=0.31\text{ GeV}
\end{equation}%
which is similar to the first soft-wall estimate (\ref{fsw1}) but
corresponds to a significantly smaller glueball mass $m_{S}^{\left( \text{sw}%
\right) }=1.37$ GeV.

The above results may be summarized as follows: (i) whereas the hard-wall
results for the ground state mass can differ by more than 30\% for Dirichlet
vs. Neumann IR\ boundary conditions, the decay constant predictions remain
in the smaller range 
\begin{equation}
f_{S}^{\left( \text{hw}\right) }\simeq 0.8-0.9\text{ GeV,}  \label{fhwf}
\end{equation}%
and (ii) the soft-wall results for the ground state decay constant center
consistently around less than half of the hard-wall value, 
\begin{equation}
f_{S}^{\left( \text{sw}\right) }\simeq 0.3\text{ GeV.}  \label{fswf}
\end{equation}%
The substantial difference between the hard- and soft-wall predictions can
be traced to the different slope of the normalized dual modes at the UV
brane (i.e. for $z=\varepsilon \rightarrow 0$). (An analogous but less
pronounced difference between the slopes of hard- and soft-wall modes was
found in the rho meson sector \cite{gri207}.) The larger slope of the
hard-wall mode translates into a larger Bethe-Salpeter amplitude at the
origin and hence into a smaller size of the scalar glueball.

In view of the sign problem which afflicts the leading nonperturbative
contributions to the soft-wall glueball correlator at distances larger than
the inverse QCD scale (cf. Sec. \ref{swcor}), and because of the exceptional
size of the missing exponential contributions, one would expect the
soft-wall results in the spin-0 glueball sector to be less reliable than
their hard-wall counterparts. This expectation is corroborated by the first
(quenched) lattice simulation of glueball decay constants \cite{che06} which
finds $f_{S}^{\left( \text{lat}\right) }=0.86\pm 0.18$ GeV. This lattice
result is inside errors fully consistent with the IOPE sum-rule value $%
f_{S}^{\left( \text{IOPE}\right) }=1.050\pm 0.1$ GeV \cite{for05}, the
instanton-liquid modes result $f_{S}^{\left( \text{ILM}\right) }=0.8$ GeV %
\cite{sch95} and our above holographic hard-wall result (\ref{fhwf}). The
soft-wall result (\ref{fswf}), on the other hand, is clearly incompatible
with the lattice prediction.

Further insight into the holographic glueball dynamics can be gained by
interpreting the above results on the basis of the structural
complementarity between the nonperturbative physics accounted for in the
soft- and hard-wall correlators (i.e. power vs. exponential contributions,
cf. Sec. \ref{gcor}). Since the large exponential contributions to the
hard-wall correlator can at least partially be associated with small-scale
instantons and are absent in the soft-wall correlator, one infers that the
instanton contribution can more than double the value of the decay constant.
The mentioned IOPE sum rule analyses \cite{for01,for05} arrived at the same
conclusion. Moreover, even the perturbative and hard instanton contributions
alone (i.e. without the unusually small power corrections and thus
analoguous to the hard-wall physics) were found to provide reasonable
approximations to the QCD $0^{++}$ glueball sum rule results for the ground
state mass and decay constant \cite{for01}. The neglect of the hard
instanton contributions, on the other hand, leads to the substantially
smaller prediction $f_{S}^{\left( \text{OPE}\right) }=0.390\pm 0.145$ GeV %
\cite{nar98} which is consistent with the soft-wall result (\ref{fswf}) but
not with the lattice value.

\section{Summary and conclusions}

\label{sum}

We have analyzed the scalar glueball dynamics contained in two approximate
holographic QCD duals, viz. the hard-wall IR brane geometry and the dilaton
soft-wall background. Our study focuses on the $0^{++}$ glueball correlation
function and its spectral density for which we have obtained closed
analytical expressions in both gravity duals. A systematic comparison with
the QCD physics content of the instanton-improved operator product
expansion, a dilatational low-energy theorem and an additional,
two-dimensional power correction then provides several new insights into the
holographic representation of hadron physics as well as estimates for
various bulk parameters of the QCD vacuum and predictions for the glueball
decay constants.

In both dual backgrounds the spectral densities are found to be
non-negative, in agreement with general principles, and to consist of an
infinite sum of zero-width glueball poles, as expected in the limit of a
large number of colors. In their representation of specific nonperturbative
glueball physics (at momenta larger than the QCD scale), however, both
holographic duals turn out to complement each other in a mutually exclusive
fashion: the soft-wall correlator contains all known types of QCD power
corrections (to leading order in the strong coupling), generated either by
condensates or by an effective UV gluon mass, while sizeable exponential
corrections as induced by small-scale instantons are found in the hard-wall
correlator. (This complementarity may in fact suggest to combine brane- and
dilaton-induced IR physics into improved QCD duals.)

As a consequence, the soft-wall correlator provides holographic estimates
for either the three lowest-dimensional gluon condensates or their Wilson
coefficients, as well as for the effective gluon mass (potentially
associated with a two-dimensional nonlocal ``condensate''), whereas the
hard-wall correlator allows for predictions of the two leading moments of
the instanton size distribution. All holographic estimates turn out to be of
the order of magnitude expected from QCD, which is at least partly a
consequence of the fact that the IR scale of both dual backgrounds is set by 
$\Lambda _{\text{QCD}}.$ The predicted signs of the two leading
dilaton-induced power corrections, however, are opposite to those of
standard QCD estimates (and in conflict with the factorization approximation
for the four-gluon condensate). We have argued that these shortcomings
provide evidence for the short-distance physics in the OPE Wilson
coefficients to be inadequately reproduced (beyond the leading conformal
logarithm) by the strongly-coupled UV regime of bottom-up models. In
conjunction with the absence of the sizeable exponential contributions, this
casts particular doubts on soft-wall results for glueball observables.

A second main objective of our analysis was to provide first holographic
estimates for the decay constants of the $0^{++}$ glueball and its
excitations, which contain valuable size information and are of direct
importance for experimental glueball searches. Our analysis shows that the
decay constants probe aspects of the dual dynamics to which the mass
spectrum is less sensitive, and thus provide a new testing ground for the
development of improved QCD duals. The hard- and soft-wall predictions for
the ground-state decay constant $f_{S}$ differ by more than a factor of two,
as do the corresponding QCD sum-rule results with and without hard instanton
contributions. In fact, as in the sum-rule analyses the enhancement of $%
f_{S} $ and the consequently reduced size of the scalar glueball\ in the
hard-wall background can be traced to the strong instanton-induced
attraction (over relatively short distances of the order of the average
instanton size) which the exponential contributions to the hard-wall
correlator generate. It is remarkable that the simple hard-wall
approximation can reproduce these small-instanton effects, which are known
to be exceptionally strong in the $0^{++}$ glueball correlator. Their
absence and the other shortcomings mentioned above render the soft-wall
predictions for the glueball decay constants unreliable, while the hard-wall
prediction $f_{S}^{\left( \text{hw}\right) }\simeq 0.8-0.9$ GeV agrees
inside errors with IOPE sum-rule and lattice results.

The above arguments for the instanton-induced origin of the decay constant
enhancement provide an example for how the complementary nonperturbative
physics in the hard- and soft-wall backgrounds, which should for the most
part generalize to other hadron channels, may be exploited to trace
differences in the holographic predictions of both backgrounds to different
origins in the soft gauge dynamics. The absence of instanton contributions
to the soft-wall correlator provides another example: since the soft-wall
dilaton background was designed to reproduce the linear trajectories of
excited mesons, it indicates that instanton effects are not directly
involved in the underlying flux-tube formation, in agreement with QCD
expectations.

Our results demonstrate that the comparison of holographic predictions with
QCD information at the correlator level can provide very specific and
quantitative insights into the gauge dynamics which different dual
backgrounds encode. This holds in particular for comparisons with the QCD
operator product expansion. Owing to its ability to factorize contributions
from short- and long-distance physics to gauge theory amplitudes, the OPE
allows for a transparent analysis and systematic improvement of several
typical shortcomings of holographic models, including those which are rooted
in their strongly coupled UV sector. These limitations notwithstanding, the
amount of glueball dynamics which we found to be represented by even the
simplest holographic duals is encouraging and indicates that the bottom-up
approach may indeed provide a viable and systematically improvable
approximation to holographic QCD.

\begin{acknowledgments}
This work was supported by the Brazilian funding agency Funda\c{c}\~{a}o de
Amparo a Pesquisa do Estado de S\~{a}o Paulo (FAPESP).
\end{acknowledgments}

\appendix

\section{Synopsis of QCD results}

\label{qcdsyn}

The instanton-improved operator product expansion (IOPE) 
\begin{equation}
\hat{\Pi}^{\left( \text{IOPE}\right) }\left( Q^{2}\right) =\hat{\Pi}^{\left( 
\text{OPE}\right) }\left( Q^{2}\right) +\hat{\Pi}^{\left( I+\bar{I}\right)
}\left( Q^{2}\right)  \label{sde}
\end{equation}%
of the scalar glueball correlator, which holds at spacelike momenta $%
Q^{2}=-q^{2}\gg \Lambda _{\text{QCD}}^{2}$, is currently known up to
operators of dimension eight, radiative corrections to the Wilson
coefficients up to $O\left( \alpha _{s}^{2}\right) $, and small-size (or
``direct'') instanton contributions of $O\left( \hbar ^{0}\right) $ to the
Wilson coefficient of the unit operator \cite{for01,for05}. The standard
part, with purely perturbative coefficients, has therefore the form (cf. %
\cite{for05,nar98} and references therein)%
\begin{align}
\hat{\Pi}^{\left( \text{OPE}\right) }(Q^{2})& =\left[ A_{0}+A_{1}\ln \left( 
\frac{Q^{2}}{\mu ^{2}}\right) +A_{2}\ln ^{2}\left( \frac{Q^{2}}{\mu ^{2}}%
\right) \right] Q^{4}\ln \left( \frac{Q^{2}}{\mu ^{2}}\right)  \notag \\
& +\left[ B_{0}+B_{1}\ln \left( \frac{Q^{2}}{\mu ^{2}}\right) \right]
\left\langle G^{2}\right\rangle +\left[ C_{0}+C_{1}\ln \left( \frac{Q^{2}}{%
\mu ^{2}}\right) \right] \frac{\left\langle gG^{3}\right\rangle }{Q^{2}}%
+D_{0}\frac{\left\langle G^{4}\right\rangle }{Q^{4}}.  \label{ope}
\end{align}%
The full set of coefficients $A_{i}$ -- $D_{i}$ can be found in Ref. \cite%
{for05}. Those needed for comparison with the holographic results below are $%
A_{0}=-\left( N_{c}^{2}-1\right) /\left( 4\pi ^{2}\right) $\ and (for the
number of colors (light flavors) $N_{c}$ $\left( N_{f}\right) =3$ \footnote{%
The $N_{f}=0$ values of the perturbative Wilson coefficients are of interest
for estimates in Yang-Mills theory without matter but should not be used
together with the phenomenologically determined values of the condensates
and instanton parameters since the latter correspond to $N_{f}=3$.}) $%
B_{0}=4+49\alpha _{s}/\left( 3\pi \right) $, $C_{0}=8$ (where a small
anomalous dimension correction has been neglected)\ and $D_{0}=8\pi \alpha
_{s}$. The gluon condensates are defined at the OPE scale $\mu $ as%
\begin{eqnarray}
\left\langle G^{2}\right\rangle &:&=\left\langle G_{\mu \nu }^{a}G^{a,\mu
\nu }\right\rangle ,\text{ \ \ \ \ \ }\left\langle gG^{3}\right\rangle
:=\langle gf_{abc}G_{\mu \nu }^{a}G_{\rho }^{b\nu }G^{c\rho \mu }\rangle , \\
\left\langle G^{4}\right\rangle &:&=14\left\langle \left( f_{abc}G_{\mu \rho
}^{b}G_{\nu }^{\rho c}\right) ^{2}\right\rangle -\left\langle \left(
f_{abc}G_{\mu \nu }^{b}G_{\rho \lambda }^{c}\right) ^{2}\right\rangle .
\label{g4fact}
\end{eqnarray}

Contributions from instantons larger than the inverse OPE scale are
accounted for in the condensates. Small-scale (or ``direct'') instantons
(and anti-instantons) contribute to the Wilson coefficients, on the other
hand, and affect dominantly the coefficient of the unit operator \cite%
{for01,for05,nov280}. In the glueball channel, the latter is given by \cite%
{for05,nov280} 
\begin{equation}
\hat{\Pi}^{\left( I+\bar{I}\right) }\left( Q^{2}\right) =\left( 4\pi \right)
^{2}\alpha _{s}^{-2}\sum_{I+\bar{I}}\int d\rho n_{\text{dir}}\left( \rho
\right) \left[ \left( Q\rho \right) ^{2}K_{2}\left( Q\rho \right) \right]
^{2}  \label{i}
\end{equation}%
($K_{2}$ is a McDonald function \cite{abr72}) where $\rho $ and $n_{\text{dir%
}}\left( \rho \right) $ denote the size and density of small instantons with 
$\rho \leq \mu ^{-1}$ in the vacuum. The nonperturbative contributions (\ref%
{i}) are known to be particularly important in the spin-0 glueball channels,
i.e. comparable to the contributions from the perturbative coefficient and
of equal or larger size than the power terms at $Q^{2}\gtrsim \Lambda _{%
\text{QCD}}^{2}$. The expression (\ref{i}) can be approximated as%
\begin{equation}
\hat{\Pi}^{\left( I+\bar{I}\right) }\left( Q^{2}\right) \simeq
2^{5}5^{2}\zeta \bar{n}\left[ \left( Q\bar{\rho}\right) ^{2}K_{2}\left( Q%
\bar{\rho}\right) \right] ^{2}  \label{ia}
\end{equation}%
where we specialized the instanton density to the spike distribution $n_{%
\text{dir}}\left( \rho \right) =\zeta \bar{n}\delta \left( \rho -\bar{\rho}%
\right) $ which becomes exact at large $N_{c}$ and where $\bar{\rho}$ and $%
\bar{n}$ are the average instanton size and density in the vacuum. The
coupling $\alpha _{s}/\pi \simeq 0.2$ is fixed at a typical instanton scale
and the factor $\zeta \simeq 0.66$ excludes contributions from instantons
with $\rho >\mu ^{-1}$ \cite{for05}.

Further QCD information on the behavior of the glueball correlator is
available in the complementary limit $Q^{2}\rightarrow 0$. Indeed, the value
of the correlator at zero momentum transfer is governed by the\ low-energy
theorem (LET) \cite{let} 
\begin{equation}
\hat{\Pi}\left( 0\right) =\frac{32\pi }{\alpha _{s}b_{0}}\left\langle
G^{2}\right\rangle +O\left( m_{q}\right)  \label{let}
\end{equation}%
where $b_{0}=11N_{c}/3-2N_{f}/3$, $m_{q}$ are the light quark masses for
flavor $q$, and UV renormalization of both sides by a dispersive subtraction
of high-frequency field contributions is implied \cite{let}. The appearance
of the gluon condensate in Eq. (\ref{let}) reflects the fact that the LET\
is a consequence of the anomalous Ward Identity for the QCD dilatation
current. Additional information on the glueball correlator has been obtained
from several versions of the instanton liquid vacuum model (ILM) in Ref. %
\cite{sch95}, whereas direct lattice information on the (point-to-point)
correlator seems currently not to exist.

\end{document}